\documentclass[runningheads]{llncs}

\usepackage[T1]{fontenc}
\usepackage{graphicx}
\usepackage{amsmath}
\usepackage{booktabs}
\usepackage{multirow}
\usepackage{xcolor}
\usepackage{pifont}

\newcommand{\cmark}{\ding{51}}%
\newcommand{\xmark}{\ding{55}}%
\newcommand{\set}[1]{\mathcal{#1}}

\usepackage{tikz}

\usepackage[normalem]{ulem} 

\usepackage{colortbl}
\usepackage{arydshln}

\newcommand{\randomids}{{\textit{track-int}}}
\newcommand{\titleids}{{\textit{artist-name-track-name}}}
\newcommand{\artisttrackids}{{\textit{artist-int-track-seq}}}

\newcommand{\dictionaryencodingcf}{{\textit{semantic-ids}}}
\newcommand{\artistiidtrackids}{{\textit{artist-iid-track-seq}}}

\newcommand{\pop}{Popularity}
\newcommand{\bm}{BM25}

\newcommand{\biencoder}{Bi-encoder$_{ft}$}
\newcommand{\biencoderzeroshot}{Bi-encoder$_{zs}$}

\newcommand{\genretrieval}{Text2Tracks}

\AtBeginDocument{%
  \providecommand\BibTeX{{%
    \normalfont B\kern-0.5em{\scshape i\kern-0.25em b}\kern-0.8em\TeX}}}

\newcommand\blfootnote[1]{%
  \begingroup
  \renewcommand\thefootnote{}\footnote{#1}%
  \addtocounter{footnote}{-1}%
  \endgroup
}

\begin{document}

\title{\genretrieval: Prompt-based Music Recommendation via Generative Retrieval}

\author{
    Enrico Palumbo\textsuperscript{$\ddagger$} \and 
    Gustavo Penha\textsuperscript{$\ddagger$} \and 
    Andreas Damianou \and 
    José Luis Redondo García\and 
    Timothy Christopher Heath\and 
    Alice Wang \and 
    Hugues Bouchard \and 
    Mounia Lalmas
}

\institute{Spotify\\}

\authorrunning{Palumbo, Penha, et al.}

\maketitle

\blfootnote{$^{\ddagger}$ Both authors contributed equally to this research.}

\begin{abstract}
In recent years, Large Language Models (LLMs) have enabled users to provide highly specific music recommendation requests using natural language prompts (e.g. \textit{``Can you recommend some old classics for slow dancing?''}).
In this setup, the recommended tracks are predicted by the LLM in an autoregressive way, i.e. the LLM generates the track titles one token at a time. While intuitive, this approach has several limitation. First, it is based on a general purpose tokenization that is  optimized for words rather than for track titles. Second, it necessitates an additional entity resolution layer that matches the track title to the actual track identifier. Third, the number of decoding steps scales linearly with the length of the track title, slowing down inference.  In this paper, we propose to address the task of prompt-based music recommendation as a generative retrieval task. Within this setting, we introduce novel, effective, and efficient representations of track identifiers that significantly outperform commonly used strategies. We introduce \emph{\genretrieval{}}, a generative retrieval model that learns a mapping from a user's music recommendation prompt to the relevant track IDs directly. Through an offline evaluation on a dataset of playlists with language inputs, we find that (1) the strategy to create IDs for music tracks is the most important factor for the effectiveness of \emph{\genretrieval{}} and semantic IDs significantly outperform commonly used strategies that rely on song titles as identifiers (2) provided with the right choice of track identifiers, \emph{\genretrieval{}} outperforms sparse and dense retrieval solutions trained to retrieve tracks from language prompts.

\end{abstract}

\section{Introduction}

\begin{figure}[h]
    \centering
    \includegraphics[width=\textwidth]{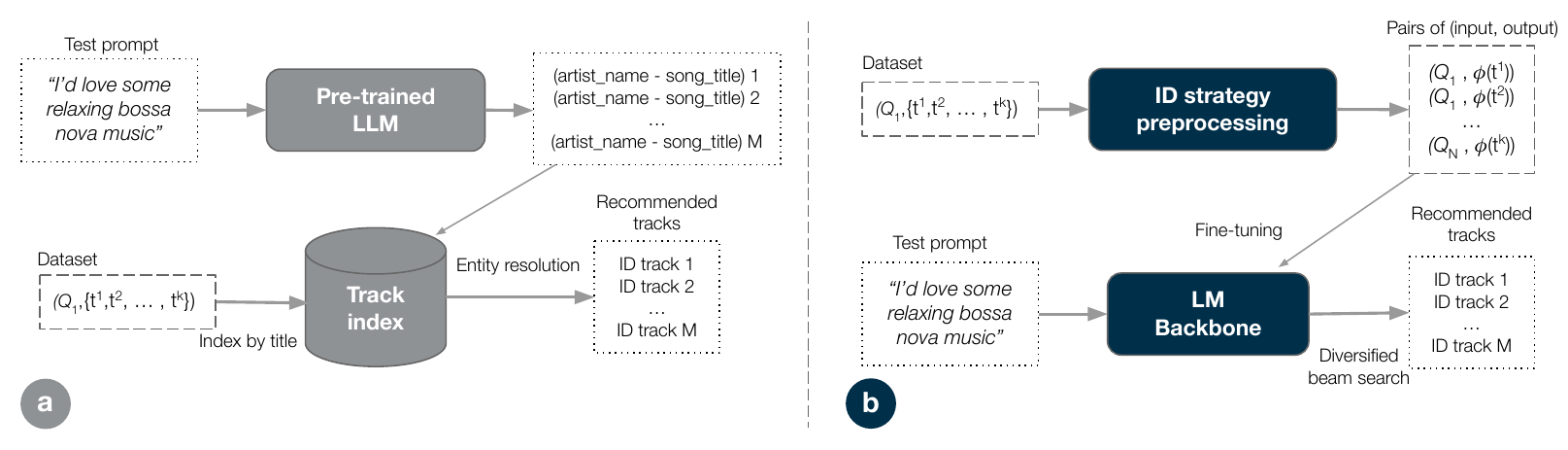}
    \caption{(a) Pre-trained LLMs deal with prompt-based music recommendation by generating the recommended artist name and song title, which are then resolved against an index to find the actual track identifiers.
    (b) \emph{\genretrieval{}} is a generative track retrieval model composed of a component that represents tracks, i.e. the ID strategy $\phi$ that maps from a track to its ID, and a backbone LM that is fine-tuned with pairs of music recommendation queries and track IDs. At test time \emph{\genretrieval{}} generates a set of recommended tracks using a diversified beam search strategy.}
    \label{fig:intro_diagram}
\end{figure}

Conversational assistants such as ChatGPT~\cite{ouyang2022training} are getting increasingly more popular thanks to recent advances in Large Language Models (LLMs)~\cite{achiam2023gpt,touvron2023llama,fitzgerald2022alexa}. A prominent feature of modern LLMs is the ``\textit{knowledge}'' stored in their weights about entities such as songs, films, or books~\cite{penha2020does}. Being able to connect media entities to their natural language descriptions is extremely useful in cold-start~\cite{sanner2023large} or conversational recommendation scenarios~\cite{deldjoo2024review}, where user preferences are often elicited through natural language prompts. Conversational Recommenders (CR)~\cite{zamani2023conversational,deldjoo2021towards} have been proposed as a powerful paradigm to elicit user preferences via language prompts, enabling complex and specific content requests, multi-turn refinements, and explanations. Typically, CRs have been implemented as systems comprising many components~\cite{jannach2021survey,chaganty2023beyond} such as query understanding, item retrieval/recommendation, dialogue management, and response generation. However, modern autoregressive LLMs can potentially do multiple of these steps in a zero-shot way by simply predicting the next token~\cite{he2023large}. 

In this context, the track recommendation step, where the LLM generates track identifiers based on a language prompt, can be seen as a Generative Retrieval problem (GR)~\cite{tay2022transformer}. GR has shown promising results for question answering and document retrieval tasks~\cite{tay2022transformer,nci}. Unlike sparse and dense approaches, GR does not rely on pre-computed indexes for the documents. In GR, a transformer model stores information about all the documents in the catalog within its parameters and directly generates document IDs for an input query. GR is particularly appealing in a conversational interface, as it opens the door to a single fine-tuned LLM handling all aspects of conversational recommendations, generating item identifiers in the same way it generates other textual content, such as follow-up questions, and explanations.

A critical question in the GR scenario is how to best represent item identifiers. In the music domain, pre-trained LLMs handle this step by generating the artist name and song title in plain text, as this is the most common representation found in their pre-training web data (Figure~\ref{fig:intro_diagram}a). However, modeling tracks through their names has several shortcomings. First, artist and track names do not generally convey meaningful semantic information about the underlying items.
Second, given the same artist and song name, finding the corresponding track identifier to play requires an additional entity resolution step. 
Third, representing track items by their names requires a number of decoding steps that are proportional to their length.

In this paper, we model the prompt-based music recommendation task, where user preferences are expressed through broad-intent natural language queries (e.g. \textit{``Can you recommend some upbeat rock tracks to dance to?''}), as a generative retrieval problem. We propose an approach to generative track retrieval called \emph{\genretrieval{}} (Figure~\ref{fig:intro_diagram}b). We focus on the problem of finding effective and efficient identifiers for music items for a generative retrieval model in the context of a prompt-based music recommendation system.  To learn how queries relate to tracks, \emph{\genretrieval{}} leverages a language model backbone that maps from queries to track IDs. 

We perform experiments on a playlist dataset for prompt-based music recommendation. Our findings reveal that (1) leveraging an ID strategy that learns semantic IDs from collaborative filtering embeddings is the most effective approach, outperforming the commonly used artist name and track title approach by 48\% in Hits@10 while reducing the number of decoding steps by $\sim$ 7.5 times; and (2) the generative retrieval approach \emph{\genretrieval{}} outperforms sparse and dense retrieval baselines.

\section{Related Work}

\paragraph{Prompt-based Recommendation}
Language-based preferences and textual inputs have always been a core part of recommender systems. Content-based recommender systems~\cite{lops2011content} have been around for several decades, matching user profiles with item textual metadata such as descriptions, reviews, etc. LLMs have further increased the attention toward the elicitation of language-based preferences for items, thanks to their ability to understand complex requests that rival traditional item-based recommenders, especially in cold-start scenarios~\cite{sanner2023large}.
Conversational recommendation is a type of conversational information-seeking activity~\cite{zamani2023conversational} and is defined as \emph{``a software
system that supports its users in achieving recommendation-related goals through a multi-turn dialogue''}~\cite{jannach2021survey}.  Conversational recommenders have the potential to achieve multiple goals such as better eliciting the user information need, recommending certain items, explaining the recommendations, answering questions about the items, etc. 

CRs have been typically implemented by pipelines with many components~\cite{jannach2021survey,chaganty2023beyond} such as query understanding, item retrieval \& recommendation, dialogue management, and response generation. With the breakthroughs of LLMs that can potentially do all those tasks, end-to-end approaches have become more popular~\cite{li2023conversation,wang2023rethinking,he2023large}. One particularly challenging aspect of conversational recommendation with LLMs is item retrieval, where broad descriptions must be matched against the item collection. A common strategy in generative models is to predict the title of the item in an autoregressive fashion~\cite{he2023large,de2020autoregressive}, and match the generated titles against the item collection afterward. Even though this strategy works with API-based LLMs, more effective and efficient solutions can be achieved when fine-tuning a model for the prompt-based recommendation task. While research in this space is progressing at unprecedented speed, none of these works models the prompt-based music recommendation task as a generative retrieval problem.

\paragraph{Generative Retrieval} 
Traditional retrieval systems rely on keyword matching to identify relevant documents for a given query~\cite{robertson2009probabilistic}. Documents are not matched if they use different words to express the same concept~\cite{van2017remedies,ji2014information}. Approaches to improve semantic matching between queries and documents have largely benefited~\cite{yates2021pretrained} from breakthroughs in natural language processing achieved by transformer-based models~\cite{vaswani2017attention}, such as BERT~\cite{devlin2018bert} and T5~\cite{t5}. For the retrieval task, transformer-based models have been used to augment documents~\cite{nogueira2019doc2query} and queries~\cite{mao2020generation}, learn representations for sparse retrieval methods~\cite{nguyen2023unified,formal2021splade}, and used as backbone for Bi-encoders~\cite{reimers2019sentence,izacard2021unsupervised,karpukhin2020dense} and Cross-encoders~\cite{monot5}.

Generative retrieval has emerged as a new promising paradigm for semantic search. In GR, transformers act as differentiable search indexes~\cite{dsi,nci,bevilacqua2022autoregressive,chen2023understanding}, learning to generate relevant document identifiers (IDs) for a specific query. Because of the output space, all documents in the collection need to be part of the training set of GR models, as their IDs need to be learned and stored in the model weights, leading to challenges of scalability~\cite{pradeep2023does} and ingestion of new documents~\cite{kishore2023incdsi}. A key point of GR models is assigning IDs for each document in the collection~\cite{tay2022transformer,hua2023index,li2023multiview,rajput2023recommender} as adding one new token to the vocabulary representing each document in the collection becomes intractable with reasonably sized collections. 

\section{Method}
We start by formally defining prompt-based music recommendation as a generative track retrieval task, followed by the main components of the \emph{\genretrieval{}} model. Figure~\ref{fig:intro_diagram}b displays a diagram of the model, showcasing how the training dataset $\set{D}$ is first pre-processed into pairs of training instances using an ID strategy $\phi$ to represent the items. Then, the Language Model (LM) is fine-tuned and at test time uses diversified beam search to generate track recommendations. We first describe how the ID strategies work before introducing the LM backbone.

\paragraph{Prompt-based Music Recommendation} The modeling assumption of this paper is that prompt-based music recommendation can be framed as a generative track retrieval task. 
The track retrieval task is defined as retrieving a set of tracks relevant to a given music recommendation query. Formally let $\set{D}=\{(Q_i, \{t^1, t^2, ..., t^{k}\})\}_{i=1}^{N}$ be a dataset composed of relevance labels for music recommendation queries, where $Q$ is the query containing the music recommendation information need and $\{t^1, t^2, ..., t^{k}\}$ are the tracks that are relevant for this query. For the conversational recommendation setting, $Q^{\tau} = \{u_0, ..., u_\tau\}$ is the set of utterances from the user until the turn $\tau$ of the dialogue. The task is then to learn a function $f(Q)$ that maps from a query $Q$ to a subset of the entire collection of tracks $\set{T}$ in a retrieval-like fashion:  $f(Q) \rightarrow $ $\{t^1, t^2, ..., t^{m}\}$, where $m \ll |\set{T}|$. Following the GR approach~\cite{tay2022transformer}, we pose that $f(Q)$ is a transformer model with a decoder layer that directly generates the subset of relevant track IDs (see Section~\ref{section:lmbackbone}). In this setup, a crucial question is how to model track IDs.

\paragraph{\textbf{\genretrieval{}: ID strategies}}
\label{sec:id_strategies}

\begin{figure}[]
    \centering
    \includegraphics[width=\textwidth]{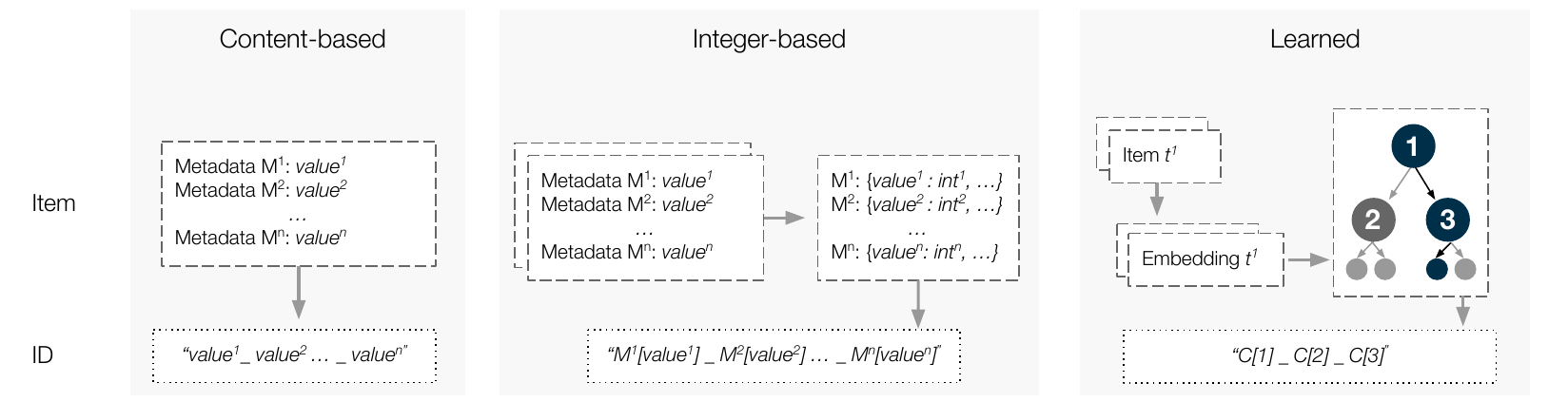}
    \caption{The three categories of ID strategies using ``\_'' as a separator. Content-based strategies use textual metadata associated with the item. Integer-based approaches use random integer values for each metadata, potentially leveraging the hierarchy of metadata available. Learned approaches go from embeddings that represent the item to hierarchically structured tokens.}
    \label{fig:id_strategies_diagram}
\end{figure}

The ID strategy is responsible for generating a string identifier for each item in the collection $\set{T}$. We explore three types of IDs: based on the content of the item, based on integers, and learned ones. Figure~\ref{fig:id_strategies_diagram} describes the three classes of approaches at a high level.

\subsubsection{Content-based} This category of IDs uses the textual content of the item as an identifier. The advantage of this type of strategy is that the knowledge stored in the weights of the LM backbone model from its pre-training procedure can be leveraged, as the text metadata is also in the space of natural language. For a single item, each metadata category $M$ has a value. For the tracks $t_i$, we have the artist name ($M^{1}$) and track title ($M^{2}$) metadata categories which leads to the \titleids{} approach: $\phi(t_i) = concat(value^{M1(t_i)}, value^{M2(t_i)})$. Table~\ref{table:id_strategy_examples} shows an example of an item and its respective \titleids{} representation, where each value of the available metadata is concatenated.

\subsubsection{Integer-based} This category of IDs uses integer identifiers to represent the items in the catalog. The most naive one is to assign random integers for each item in the collection: \randomids{}. When using this strategy we do not add the IDs to the vocabulary of the LM backbone model, as adding millions of tokens to the vocabulary does not scale practically. This means that the LM backbone model will split the IDs into several tokens, and needs to learn to output tokens that when combined represent an item. 

To leverage all available metadata we propose to have separate integers to represent each metadata value, which allows for tokens to be shared across items with the same metadata values. Given that the artist and track metadata categories are hierarchical, i.e. tracks belong to a main artist, we first generate random integers for the artist metadata values ($M^1$) and then we sequentially count in order of appearance in the training dataset the track name metadata values ($M^2$) of that artist.  So for example, if the artist $A1$ has three different tracks $t^1,t^2,t^3$ they would be represented as follows: ``\textit{1\_1}'', ``\textit{1\_2}'', and ``\textit{1\_3}''\footnote{Following~\cite{hua2023index} we start the counts for artists and tracks within that artist from 1000.}. We refer to this strategy as \artisttrackids{}.

To facilitate learning a representation for certain items, we can also add a limited number of tokens related to a metadata category to the vocabulary. In \artistiidtrackids{} we add the top-K popular artist's integers in terms of appearance in the training set. This allows the model to represent such artists with a single token instead of the combination of the tokens that the tokenizer would otherwise generate. The remainder of the artists (not in the top-K) get their tracks represented as in the previous approach \artisttrackids{}.

\subsubsection{Learned} This category of IDs learns how to discretize the embeddings that represent each item in the collection. Ideally, similar items in the embedding space share more tokens after this discretization is done by the learned approach. So for example, if two tracks $t^1, t^2$ share the same genre, a learned approach could create the following hierarchy in their tokens, by using a shared token ``\textit{<0>}'' in the beginning: ``\textit{<0><1>}'' and ``\textit{<0><2>}''.  Learned strategies have also the benefit of controlling how many tokens each track is composed of, as well as the number of tracks that have the same tokens. While the other approaches in the paper uniquely identify each item with an ID, learned strategies can hash multiple tracks with the same tokens,\footnote{If the predicted ID has multiple tracks associated with it we sort by popularity.} which allows for compression.

We test two different ways to encode each item in an embedding space that is used by the learned ID strategy afterward. The first approach is \textit{text-based}. For each track, it creates a textual representation for it by concatenating the train set queries where the track appears. To go from the text to an embedding it uses a sentence encoder model. 

The second approach is \textit{cf-based}, and it relies on collaborative filtering track embeddings built by leveraging co-occurrence track patterns in playlist data.

To map vectors into semantic IDs, we use a discretization algorithm inspired by the use of sparse coding for word sense induction~\cite{arora2018linear}. We learn a sparse coding~\cite{olshausen1997sparse} for each item, which is an approximation of the item $embedding(t)$ as a sparse linear combination of a much smaller set of unit vectors, called the dictionary. After specifying the size of the dictionary, $s$, and the number of non-zero coefficients in each coding, $c$, we use a standard dictionary learning algorithm~\cite{mairal2009online} to simultaneously learn the vectors, $v_1 \dots v_s$, in the dictionary and the coefficients of the coding for each track, $a_1 \ldots a_s$, such that $t \approx a_1v_1 + \ldots + a_sv_s$ while keeping the magnitude of $a_1, \ldots, a_s$ small.  Ordering the non-zero coefficients from largest to smallest magnitude, $|a_{i_1}| \geq \ldots \geq |a_{i_c}|$, we obtain an ID for the track as $<\text{sgn}(a_{i_1})i_1> \ldots <\text{sgn}(a_{i_c}){i_c}>$.  This gives each track an ID with $c$ integer tokens from a lexicon of $2s$ ($s$ positively signed tokens and $s$ negatively signed ones) possible tokens with the valuable property that two tracks whose IDs start with the same subsequence of tokens are likely to have similar representations in the original embedding space.

\definecolor{Gray}{gray}{0.95}

\begin{table}[]
\caption{Examples of the IDs generated for each strategy to represent a track for \emph{\genretrieval{}}. ``<'' and ``>'' delimiters indicate that the string is added to the vocabulary of the model as a new independent token. UI indicates if the strategy uniquely identifies each track. \underline{\textcolor{teal}{Underline}} indicates track name and \dashuline{\textcolor{purple}{dashed underline}} indicates artist name.  
}
\label{table:id_strategy_examples}
\centering
\begin{tabular}{@{}lccc@{}}
\toprule
Category & ID strategy $\phi$ & UI & Example \\ \midrule 
\rowcolor{Gray}
content                   & \titleids{}       &      \cmark      & ``\textit{\dashuline{\textcolor{purple}{artist name}}\_\underline{\textcolor{teal}{track name}}}'' \\ 

\multirow{3}{*}{integer}  & \randomids{}        &   \cmark      & ``\textit{{1001}}''                      \\  
                          & \artisttrackids{}    &   \cmark     & ``\textit{\textcolor{purple}{\dashuline{1001}}\_\underline{\textcolor{teal}{1001}}}''                \\ 
                          & \artistiidtrackids{}   &   \cmark   & ``\textit{\textcolor{purple}{\dashuline{<1001>}}\_\underline{\textcolor{teal}{1001}}}''              \\
                          
  
  \rowcolor{Gray}
                          \multirow{1}{*}{learned} & \textit{semantic-ids}  &  \xmark & ``\textit{{<0><2><3>}}''               \\  
                          
                          \bottomrule
\end{tabular}
\end{table}

\paragraph{\textbf{\genretrieval{}: LM backbone}}

\label{section:lmbackbone}
We propose to fully parameterize the retrieval function $f(Q)$ with a differentiable \emph{seq2seq} transformer model. Following the intuition in~\cite{dsi}, we hypothesize that all the necessary information for the generative track retrieval can be learned and stored within the backbone language model's parameters.

\subsubsection{Training} When fine-tuning the LM backbone, each train set query $Q$ generates $k$ training instances, where $k$ is the number of relevant tracks $\{t^1, t^2, ..., t^{k}\}$ for that query. Each track is first mapped to a textual ID by one of the ID strategies, and then the model is trained with the pairs of $(Q, \phi(t))$. For the inputs where we have multiple conversational turns, we concatenate the dialogue utterances into a single query: $Q = concat(\{u_0, ..., u_\tau\})$.

\subsubsection{Inference}
At inference time, track IDs are generated via diversified beam search~\cite{vijayakumar2016diverse} for the given prompt. For each group, beam search is applied to generate track IDs; however, there is a penalty for homogeneity across the generated tokens from different groups. This means that the model is penalized (as controlled by the homogeneity hyperparameter) when the output tokens are not diverse. Diverse beam search allows \emph{\genretrieval{}} to diversify the set of predictions, which is crucial for track recommendation.

\section{Experimental Setup}
We here introduce the setup that we used to conduct our experiments.
\paragraph{Dataset} We use a dataset of playlist data. While the train and evaluation data are a subset of playlists with long descriptive titles that can be used as proxy for music recommendation prompts (e.g. \textit{chill lofi vibes to focus}), the test data is a selected set of playlists created by a team of professional music editors. The training set contains \~1M (query, track) pairs, \~75k unique queries, \~500k unique tracks. The test set contains \~2500 (query, track), 513 unique titles, 2347 unique tracks.

\paragraph{Baselines \& Implementation} Our first baseline, \pop{}, retrieves the most popular tracks regardless of the query. For the dense (Bi-encoder) and sparse (\bm{}) baselines, we represent each track by the titles of playlists they appear in the train set. So for example if $t_{i}$ appears in the playlists ``\textit{rock}'', ``\textit{metal}'' and ``\textit{guitar solos}'' we represent $t_{i}$ by their concatenation: ``\textit{rock, metal, guitar solos}''. For \bm{}~\cite{robertson1994some} we use the default hyperparameters and implementation provided by the PyTerrier toolkit~\cite{pyterrier2020ictir}. For the \biencoderzeroshot{} model, we rely on the SentenceTransformers~\cite{reimers-2019-sentence-bert} models\footnote{\url{https://www.sbert.net/docs/pretrained\_models.html}}. Specifically, we employ the pre-trained model \textit{all-mpnet-base-v2}. When fine-tuning \biencoder{} we rely on the \textit{MultipleNegativesRankingLoss}.
Generative recommendation baselines such as TIGER~\cite{rajput2023recommender} cannot be applied in this experimental setup, as they rely on a history of user interactions rather than a query with a music recommendation prompt.

\emph{\genretrieval{}} uses a T5~\cite{raffel2019exploring} base model that was instruction-tuned, known as Flan-T5~\cite{chung2022scaling}. We rely on the Hugginface library to fine-tune it, and unless otherwise stated we use the \textit{flan-t5-base} pre-trained model. We fine-tune it for a total of $20$ epochs, with a learning rate of $5e^{-4}$ and a batch size of $64$. For the ID strategy \artistiidtrackids{} we add the top $50$k artists as tokens to the vocabulary. 
For the text-based learned methods for representing IDs, we rely on the \emph{all-MiniLM-L6-v2} model from SentenceTransformers to encode the textual representation of the tracks. For the \emph{cf-based} learned methods, we rely on \emph{word2vec} model~\cite{mikolov2013distributed} to generate embeddings by training it on the sequences of tracks. For KMeans and the dictionary learning approach, we use the \textit{scikit-learn}\footnote{\url{https://scikit-learn.org/stable/}.}

We evaluate our models using the number of relevant items (Hits), with a cut-off thresholds of $10$. We use Student t-tests with Bonferonni correction and a confidence level of $0.95$ to calculate the statistical significance of different models.

\section{Results}
We now go over the results of our experiments.

\paragraph{\textbf{What is the impact of different ID strategies on \genretrieval{}'s effectiveness?}} In Table~\ref{tab:track_id_strategies} we see the results for the different strategies to represent IDs. Our results show that the ID strategy is a deciding factor for the effectiveness of \emph{\genretrieval{}}, achieving the best performance with \dictionaryencodingcf{}, showing the power of using strong collaborative filtering-based track representations and semantic IDs. The second best strategy is \artistiidtrackids{}, where tracks from the same artists share tokens, and at prediction time the model is forced to first retrieve the relevant artist, and within that artist predict the relevant track. The results show that the \textit{artist$\rightarrow$track} hierarchy provided by \artistiidtrackids{} is quite important in learning good recommendations, although it does not perform as well as best learned approaches. By comparing \artistiidtrackids{} with \artisttrackids{} we see that adding the top $50$k artists as independent tokens to the vocabulary increases the model's effectiveness.

\begin{table}[h!]
\centering
\caption{Table of track ID strategies and results. Bold denotes the highest effectiveness for the use of semantic IDs leveraging collaborative filtering embeddings.}
\renewcommand{\arraystretch}{1.2} 
\begin{tabular}{c @{\hskip 0.5cm} c @{\hskip 0.5cm} c @{\hskip 0.5cm} c} 
\hline
\textbf{Category} & \textbf{Track ID strategy}  & \textbf{Embedding} & \textbf{hits@10} \\ \hline
Content           & artist-track-name                  & -                  & 0.182            \\ \hline
\multirow{3}{*}{Integer} & track-int                           & -                  & 0.140            \\ 
                  & artist-int-track-seq                & -                  & 0.185            \\ 
                  & artist-iid-track-seq                & -                  & 0.239            \\ \hline
\multirow{2}{*}{\textbf{Learned}} & semantic-ids                         & text-based         & 0.180            \\ 
                  & \textbf{semantic-ids}               & \textbf{cf-based}  & \textbf{0.270}   \\ \hline
\end{tabular}

\label{tab:track_id_strategies}
\end{table}

The worst-performing ID representation is \randomids{} which does not add any prior knowledge to the initial representation of the tracks, and requires the model to learn how queries match groups of tokens (on average $\sim$4 tokens) that represent the tracks. When using \titleids{}, the LM can leverage the ``knowledge'' stored in its weights regarding the items. Even though the number of tokens per track is high when using \titleids{} ($\sim$15 tokens on average), it is still able to outperform \randomids{}. Overall, our best strategy \dictionaryencodingcf{} with cf-based embeddings outperforms the commonly used strategy in pre-trained LLMs \titleids{} by 48\% in Hits@10, while being much more efficient (3 decoding steps vs an average of $\sim$15 decoding steps). 

\begin{table}[]
\centering
\caption{Effectiveness of Text2Tracks and competing models. Bold denotes the highest effectiveness for \textit{hits@10}.}
\label{table:main_results}
\begin{tabular}{@{}llc@{}}
\toprule
\textbf{Model} & \textbf{How it works} & \textbf{hits@10} \\ \midrule
BM25 & Keyword indexing & 0.101 \\
Bi-encoder zero-shot & Semantic match between playlist titles & 0.065 \\
Bi-encoder fine-tuned & Fine-tuned semantic match between playlist titles & 0.119 \\
\textbf{Text2Tracks} & \textbf{GR using cf-based semantic-ids} & \textbf{0.270} \\ \bottomrule
\end{tabular}
\end{table}
\paragraph{\textbf{How does \genretrieval{} compare to baselines that retrieve tracks based on a language prompt?}} We display in Table~\ref{table:main_results} the results for the task of generative track retrieval. We see that \genretrieval{} outperforms the baselines, obtaining a 127\% increase in \textit{hits@10} with respect to the closest competitor. This proves the effectiveness of Generative Retrieval in this task, as directly fine-tuning a model to generate track identifiers significantly outperforms commonly used techniques that rely on indexing and retrieving track vectors.

We hypothesize that the gains of \emph{\genretrieval{}} over traditional sparse and dense baselines come from two main reasons. The first is the lack of textual information for the tracks in such datasets besides the track and artist names. This way all models rely exclusively on the train queries and predicting which track is relevant directly allows for more expressiveness than using similarity between the representations of the query and tracks. The second one is that \emph{\genretrieval{}} can distinguish between popular and non-popular items based on the occurrence of such tracks in the training set. While tokens representing popular artists and tracks are more likely to be generated by \emph{\genretrieval{}}, the textual representations of tracks for traditional retrieval methods cannot distinguish two tracks with different popularity values. For instance, when asked for  ``\textit{Christmas Hits}'', \emph{\genretrieval{}} reliably retrieved the most iconic holiday tracks, while other systems provided topically relevant, but less canonical entities.

\begin{figure}[]
    \centering
    \includegraphics[width=.7\textwidth]{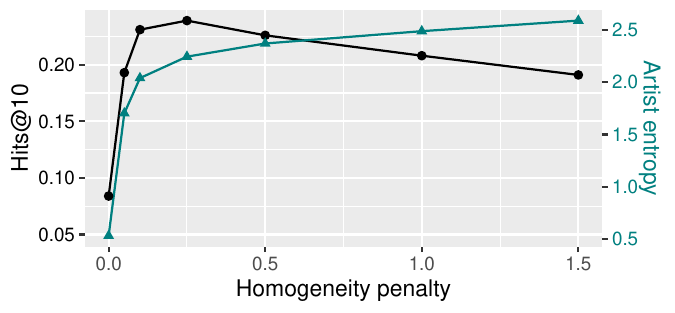}
    \caption{The effect on Hits@10 and on the diversity of the artists when increasing the homogeneity penalty hyperparameter, which applies a penalty for generating tokens that were selected in other beam search groups at prediction with \emph{\genretrieval{}} }
    \label{fig:diversity_penalty}
\end{figure}

\paragraph{\textbf{How does \emph{\genretrieval{}} diversified predictions affect effectiveness?}}
The homogeneity parameter of diverse beam search allows for control of how much the generation of similar predictions across different beam search groups is penalized~\cite{vijayakumar2016diverse}. We study the effect that this parameter has on the model's effectiveness and on the diversity of the recommended tracks, measured as the Shannon entropy of the list of predicted artists~(Figure~\ref{fig:diversity_penalty}).  We observe that the artist entropy (i.e. diversity) monotonically increases when the homogeneity penalty increases, while the effectiveness has a local maximum at $\sim$ 0.25. 

\section{Conclusions}
In this paper, we propose to address the problem of prompt-based music recommendation through the lenses of generative retrieval and we introduce a generative track retrieval model \emph{\genretrieval{}}. Given that tracks are items with scarce text, \emph{\genretrieval{}} learns to generate track IDs directly from a textual query as an end-to-end differentiable model. We propose several new strategies to represent track IDs and show that the track ID representation strategy is a crucial ingredient to obtain high effectiveness. By leveraging semantic IDs built on top of powerful collaborative filtering embeddings we obtain a 48\% increase in effectiveness with respect to the commonly used strategy that model tracks through their title. We then show that for \emph{\genretrieval{}} outperforms dense and sparse retrieval solutions with a large margin of 127\%, proving the effectiveness of the generative retrieval approach in this task.  These findings show that the problem framing that we propose combined with the novel ID strategies that we introduce for music tracks lead to sizable improvements both in terms of effectiveness and efficiency of prompt-based music recommendation. 

We believe that this work represents a stepping stone in the use of generative retrieval in the music domain, leading to more accurate models that can also greatly simplify based on indexing and retrieving vectors. The generative retrieval approach is particularly appealing in a conversational setup as it paves the way to a fully fine-tuned conversational recommender that can generate track IDs jointly with explanations, follow-up requests, and any other textual content. In future work, we plan to extend this approach to generate joint track recommendations and textual responses, experimenting with decoder-only models such as Llama~\cite{touvron2023llama}. We also aim to test the generalizability of our findings to other media items such as podcasts or books.


\bibliographystyle{splncs04}
\bibliography{biblio}


\end{document}